\documentclass[a4paper,11pt]{article}

\usepackage{pos}
\usepackage{lipsum} 
\usepackage{wrapfig}
\usepackage{graphicx}
\usepackage{caption}  
\usepackage{subcaption} 
\usepackage{lineno}
\usepackage{siunitx}
\usepackage{graphicx}
\usepackage{subcaption}
\usepackage{booktabs}

\title{ Impact of Hole-ice Calibration on High Energy Event Reconstruction with the IceCube Upgrade }

\author{The IceCube Collaboration \\{\normalsize \normalfont(a complete list of authors can be found at the end of the proceedings)}\\}

\emailAdd{kdutta@icecube.wisc.edu}
\emailAdd{tyuan@icecube.wisc.edu}
\emailAdd{steve.sclafani@icecube.wisc.edu}
\emailAdd{sarah.mancina@icecube.wisc.edu}

\abstract{
The IceCube Upgrade, currently under construction at the geographic South Pole, is the next development stage of the IceCube detector. It will consist of seven new columns of novel optical sensors and advanced calibration devices densely deployed at the center of the existing array. The sensors are frozen into the ice in boreholes created by hot water drilling. The refreezing forms hole ice around modules with optical properties that differ from the surrounding glacial ice. A key objective of the IceCube Upgrade is to enhance our understanding of the optical properties of both bulk ice and refrozen hole ice. Precise ice modeling is crucial for the directional reconstruction of TeV-PeV neutrinos, as resolutions at such high energies can be strongly impacted by uncertainties in ice properties and optical sensor response. An improved directional reconstruction performance will translate to a boost in neutrino source sensitivities using IceCube data collected over the last 12 years. In this contribution, we present the expected improvements in reconstruction performance resulting from advances in hole-ice modelling and the resulting impact on IceCube's sensitivity to astrophysical neutrino sources across three distinct event samples.

\vspace{4mm}

{\bfseries Corresponding authors:}
Kaustav Dutta$^{1*}$, 
Tianlu Yuan$^{2}$, 
Steve Sclafani$^{3}$,
Sarah Mancina$^{4}$\\
{$^{1}$ \itshape Johannes Gutenberg Universität, Mainz, Germany}\\
{$^{2}$ \itshape University of Wisconsin-Madison, 500 Lincoln Dr, Madison, WI 53706, USA}\\
{$^{3}$ \itshape University of Maryland, College Park, MD 20742, USA}\\
{$^{4}$ \itshape University of Padova, Via VIII Febbraio, 2, 35122 Padova PD, Italy}\\[4mm]

$^*$ Presenter
}

\ConferenceLogo{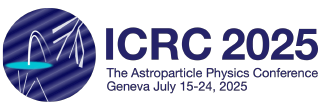}

\FullConference{39th International Cosmic Ray Conference (ICRC2025)\\
 15–24 July 2025\\
Geneva, Switzerland\\}

\let\oldthebibliography\thebibliography
\let\endoldthebibliography\endthebibliography
\renewenvironment{thebibliography}[1]{%
  \oldthebibliography{#1}%
  \setlength{\itemsep}{0pt}
  \setlength{\parskip}{0pt}
}{%
  \endoldthebibliography}%

\begin{document}

\maketitle

\section{Hole ice calibration with the IceCube Upgrade}\label{sec1}

\indent The IceCube Neutrino Observatory is a cubic-kilometer-scale neutrino detector situated at the geographic South Pole~\cite{Instrumentation}. The next stage of the project, known as the IceCube Upgrade~\cite{Ishihara_upgrade}, is currently in its deployment phase and is expected to be completed by the austral summer of 2026. The Upgrade will deploy seven densely instrumented strings with a reduced horizontal and vertical spacings of \SI{20}{\meter} and \SI{3}{\meter}, respectively, compared to the current \SI{125}{\meter} and \SI{17}{\meter} in the sparser array and \SI{70}{\meter} and \SI{7}{\meter} in the DeepCore \cite{DeepCore} region. The modules will be deployed in the clearest region of Antarctic ice, between 2150 and \SI{2425}{\meter} in depth. New optical sensor designs \cite{mDOM,DEgg} will be introduced that will improve reconstruction at low energies and enhance the calibration of both the detector geometry and optical properties of ice.\\
\indent For reconstructing low-energy events in the GeV range, the primary limitation arises from the low number of detected photons, leading to significant statistical uncertainties that outweigh systematic effects such as photon scattering in the ice \cite{Ishihara_upgrade}. At higher TeV-PeV energies, the increased number of observed scattered photons reduces the statistical uncertainty and the reconstruction accuracy is increasingly limited by systematic uncertainties, particularly those related to light propagation through the ice. These include both the geologically formed glacial bulk ice and the local ice surrounding each optical module, the latter of which is the focus of this study.\\
\begin{figure}[htbp]
    \centering
    \begin{subfigure}[t]{0.48\textwidth}
        \centering
        \includegraphics[width=\textwidth]{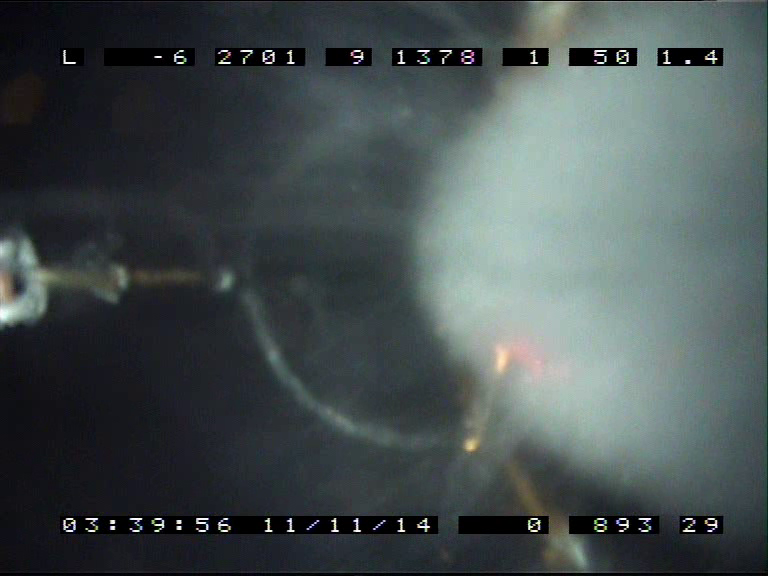}
    \end{subfigure}
    \hfill
    \begin{subfigure}[t]{0.48\textwidth}
        \centering
        \includegraphics[width=\textwidth]{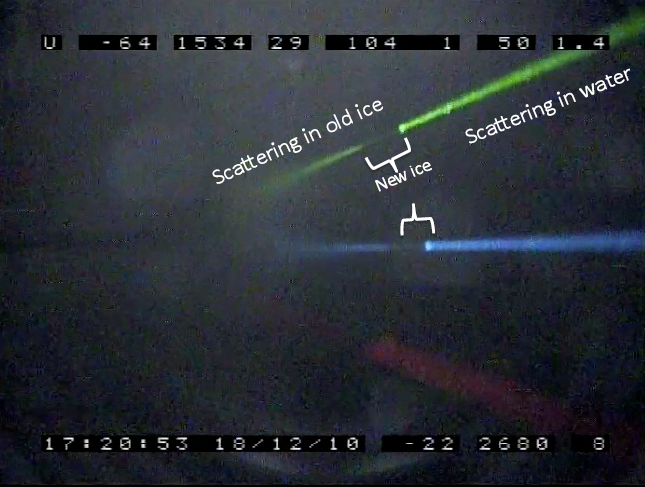}
    \end{subfigure}
    \caption{Images from integrated cameras shortly after deployment. Left: Upward-facing camera captures a milky, bubble-rich region on the right in contrast to the clear refrozen ice on the left. Right: Side-facing view shows significantly reduced scattering in the refrozen ice compared to the surrounding glacial ice.}
    \label{Sweden_Camera}
\end{figure}
\indent Each IceCube deployment hole is created using hot-water drilling. As the melted ice refreezes, it forms a vertical column that is optically clearer than the surrounding bulk glacial ice. The drilling process also introduces bubbles and impurities which are displaced toward the center of the hole or stick inhomogeneously to the glass housing of optical modules \cite{hole_ice_1}. Camera footage (Fig.~\ref{Sweden_Camera}) and in-situ calibration data~\cite{led_paper} have shown that this central "bubble column" region exhibits significantly shorter scattering and absorption lengths compared to the surrounding ice.  In IceCube terminology, “hole ice” specifically refers to this bubble-rich core because this is the region that affects detector sensitivity. Since all photons travel through this region before reaching a PMT, precise modeling of the hole ice is essential to improve high-energy event reconstruction and overall detector performance.

\section{Event selections}
\label{event_selections}
\indent The Monte Carlo (MC) event simulation includes photon propagation through ice and subsequent detector response to the observed photons. These simulations are seeded using actual event selection datasets, which include the neutrino interaction vertex, energy, and arrival directions in zenith and azimuth. The study uses three different datasets, each containing a high-purity sample of either cascades or tracks, as discussed below.

\subsection{High Energy Starting Event Sample (HESE)}
\indent The HESE selection \cite{HESE} identifies neutrino interactions that start within a defined fiducial volume of the IceCube detector by using the outermost layers of Digital Optical Modules (DOMs) as a veto against incoming atmospheric muons and neutrinos. Selected events are required to deposit at least 6000 photoelectrons (PE) within the fiducial volume and have a reconstructed energy greater than 10 TeV. The sample includes all neutrino flavors and topologies but in this study, only $\nu_e$ and $\bar{\nu_e}$ events have been considered. The expected event population includes a diffuse flux of astrophysical neutrinos at TeV-PeV energies, with contributions from both northern and southern sky directions.  

\subsection{Deep Neural Network (DNN) Cascade Selection }
\indent The DNN Cascade selection \cite{DNN_cascade} in IceCube targets high-purity samples of cascade-like neutrino events over a broader energy range between 1~TeV and several PeV. It employs convolutional neural networks (CNNs) trained on charge and time distributions recorded by the modules, to effectively classify event topologies and lower the background rate. The selected cascade events result from neutral-current interactions of all flavors and charged-current interactions of electron and tau neutrinos. The resulting dataset is well-suited for identifying neutrino sources in the Southern Sky, particularly diffuse emission from the Galactic Plane.

\subsection{Enhanced Starting Track Event Selection (ESTES)}

\indent ESTES \cite{ESTES} is an event selection designed to identify track events that start within the detector. Because the angular resolution of high-energy track events is better  than that of cascades (< 1° at 100 TeV \cite{ESTES}), the selection employs a dynamic veto technique which reconstructs the interaction vertex and calculates the probability that the event started within the detector by comparing the expected photon yields in the modules from incoming muons before and after the reconstructed vertex with the observed signals. This technique reduces the atmospheric muon background to less than one expected event per year. The resulting high-purity sample allows for astrophysical neutrino detection from the southern sky above 10 TeV, supporting both diffuse and point-source searches.

\section{Simulation \& Reconstruction}\label{sec2}
\label{sim_reco}
\indent The simulation setup includes the full IceCube detector geometry. The simulation incorporates state-of-the-art knowledge of glacial ice, including depth-dependent optical properties \cite{led_paper}, anisotropic scattering due to birefringence \cite{birefringence}, and azimuthal layer tilts \cite{undulations}. Each photon is tracked through successive scatterings, where the scattering angle is sampled from an appropriate phase function. If the photon reaches a DOM or if the propagated length exceeds the absorption length, it is taken out of the photon pool \cite{led_paper}. The detector response is simulated by incorporating the angular acceptance of the DOM, PMT quantum efficiency, conversion of photons to photoelectrons (PE) inside the PMT, and the digitization of the output signal by the DOM electronics. A high-statistics MC dataset is used to determine the expected PE yield at each DOM as a function of distance and angle relative to the event. These distributions are then fitted with multi-dimensional spline functions \cite{splines}.
\begin{wrapfigure}[17]{l}{0.48\textwidth}
\vspace{-4mm}
    \centering
    \includegraphics[width=\linewidth]{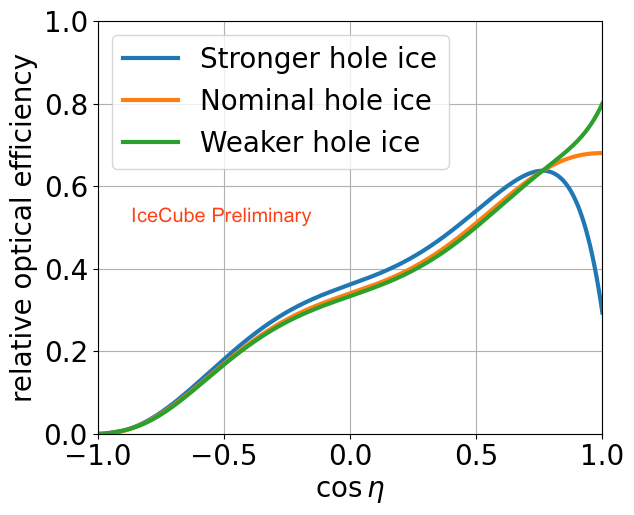}
    \caption{Effective angular acceptance curves representing the nominal, stronger, and weaker hole-ice scenarios. Each curve is normalized to preserve the total detection efficiency.}
    \label{fig:angular_acceptance_hole_ice}
\end{wrapfigure}
\indent Figure~\ref{fig:angular_acceptance_hole_ice} shows the effective angular acceptance curves \cite{led_paper} under three hole ice scenarios, as a function of the incident angle $\eta$ of photons relative to the PMT axis. Here, $\cos\eta = 1$, 0, and –1 correspond to photons arriving directly from the front, side, and behind the PMT, respectively. Due to the bubble column, photons approaching from the front are absorbed before reaching the module. This results in a decreased module sensitivity at the front of the PMT compared to laboratory measurements. Therefore, hole ice effects are typically incorporated by reducing the forward sensitivity in the angular response curve, with suppression proportional to the impact of hole ice \cite{led_paper}. The term "Nominal" refers to the scenario in which event reconstruction is performed using the true, simulation-derived spline tables that represent the correct PE expectations for the true hypothesis. "Stronger" and "Weaker" cases represent the extreme boundaries of the hole-ice parameter space constrained by flasher data \cite{led_paper}. The parameterizations (table~\ref{table_hole_ice_wrap}) are based on a model \cite{MSU_model} that treats the forward sensitivity as a free parameter. Re-simulations with different hole-ice models mimic a parameter knowledge bias which offsets the calculated PE yields from those derived from the true splines. This leads to a deterioration in the directional reconstruction of an event.
\begin{wraptable}[11]{r}{0.28\textwidth}
\vspace{-4mm}
\centering
\begin{tabular}{|l|c|c|}
    \hline
    \textbf{Model} & \textbf{p} & \textbf{p2} \\
    \hline
    Nominal  & 0.35 & 0  \\
    Stronger & 0.35 & -3 \\
    Weaker   & 0.35 & 1  \\
    \hline
\end{tabular}
\caption{Parameters defining the angular acceptance curves shown in Fig.~\ref{fig:angular_acceptance_hole_ice}. The values of p and p2 modify the mid-angle and forward-angle regions of the sensitivity curve.} 
\label{table_hole_ice_wrap}
\end{wraptable}

\indent For cascade reconstruction in IceCube, events are usually approximated as point-like light sources \cite{point_like} because of the negligible longitudinal length of the cascade compared with the distances between modules.  However, previous studies \cite{double_cascade} have shown that at high energies the longitudinal extent of the shower becomes significant. Modeling this extent with a two-cascade approach improves reconstruction accuracy. The reconstruction procedure begins with a standard single-cascade fit to obtain an initial estimate of the parameters, which are then used as a seed for the two-cascade fit. For high-energy muon tracks, the Millipede algorithm \cite{Millepede} accounts for stochastic energy losses by segmenting the track into cascade-like light sources caused by processes such as bremsstrahlung and photo-nuclear interactions. A maximum likelihood fit to photon arrival times in hit modules yields the best-fit muon direction.\\

\section{Impact of hole-ice parameterizations}\label{sec3}
Previous studies \cite{Henningsen} using systematic fits to MC data from light emitting sources simulated on Upgrade strings showed that the Upgrade can achieve an excellent precision of $\sim$1\% in the forward region (cos $\eta$ > 0.5) and $\sim$0.1\% in the middle region (cos $\eta$ $\approx$ 0) of the modified angular acceptance curves. This is about an order of magnitude better than the constraints achieved from nuisance parameter fits in DeepCore oscillation analyses \cite{Hole_ice_dc_osc}. Although simulation-based projections tend to be more optimistic than measured data, an improvement in the calibration precision is still expected from the Upgrade. We now explore how the improved modeling of hole ice impacts the directional reconstruction performance for high-energy events and its contribution to IceCube sensitivities.

\subsection{Reconstruction resolutions}
\begin{figure}[htbp]
    \centering
    \begin{subfigure}[t]{0.48\textwidth}
        \centering
        \includegraphics[width=\textwidth]{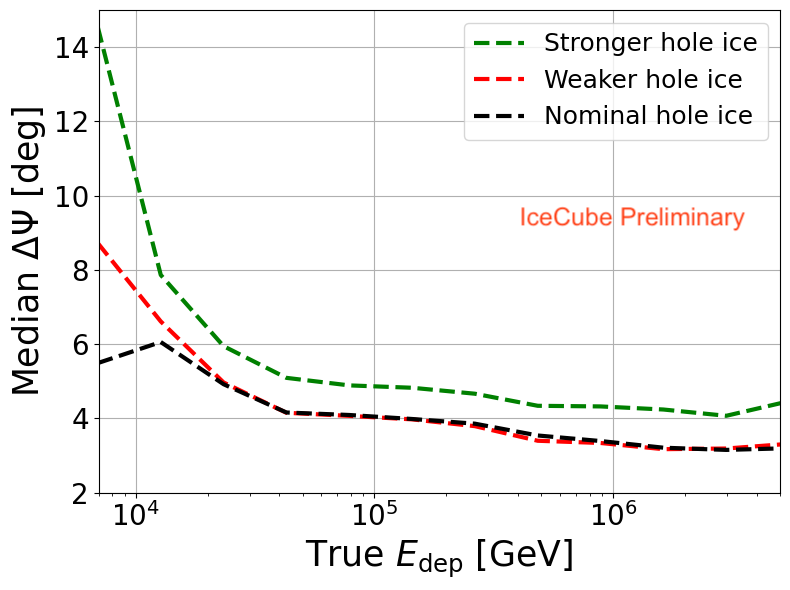}
    \end{subfigure}
    \hfill
    \begin{subfigure}[t]{0.48\textwidth}
        \centering
        \includegraphics[width=\textwidth]{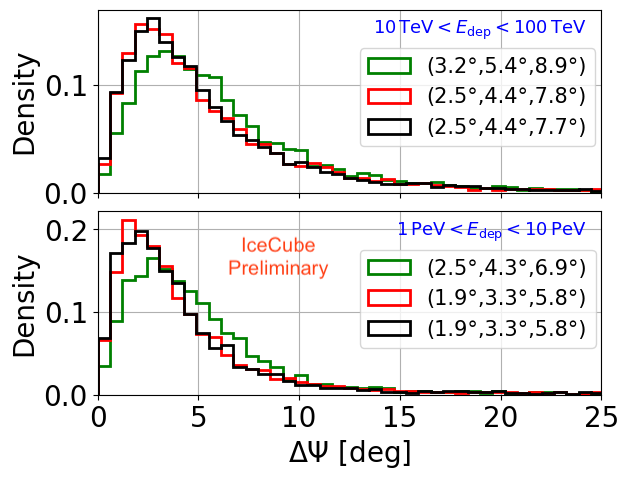}
    \end{subfigure}
    \caption{Left: Reconstruction resolution $\Delta\Psi$ on HESE MC under nominal, weaker, and stronger hole-ice assumptions, shown as a function of deposited cascade energy $E_{\text{dep}}$. Right: Distributions of $\Delta\Psi$ for the three scenarios in two energy slices. The legend indicates the (25\%, 50\%, 75\%) quantile values.}
    \label{HESE_resolutions_forward}
\end{figure}

\indent The angular resolutions under nominal, weaker, and stronger hole-ice assumptions, relative to the HESE MC benchmark are shown in the left panel of Fig.~\ref{HESE_resolutions_forward}. A weaker hole-ice assumption leads to a degradation of about 25\% at 10 TeV which becomes negligible at higher energies. A stronger hole-ice assumption initially worsens the resolution by over 100\% at 10 TeV which stabilizes at $\sim$ 33\% beyond 100 TeV. The resolution uncertainties (Fig.~\ref{HESE_resolutions_forward}, right panel) show a slight increase for the stronger-hole ice case but a minimum impact for the weaker case. Modifying the forward angular acceptance region was also found to be more significant in resolution gains compared to changing the intermediate region. The results suggest that improvements in hole-ice calibration from the Upgrade are expected to have only a minor effect on the reconstruction performance of contained high-energy cascades. Given the comparatively smaller influence of the intermediate region, the remainder of this study will focus on variations in the forward angular acceptance.

\begin{figure}[htbp]
    \centering
    \begin{subfigure}[t]{0.47\textwidth}
        \centering
        \includegraphics[width=\textwidth]{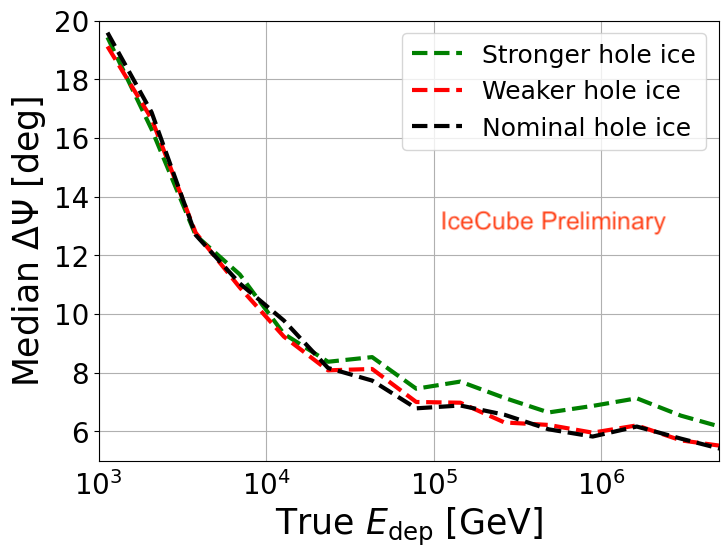}
    \end{subfigure}
    \hfill
    \begin{subfigure}[t]{0.48\textwidth}
        \centering
        \includegraphics[width=\textwidth]{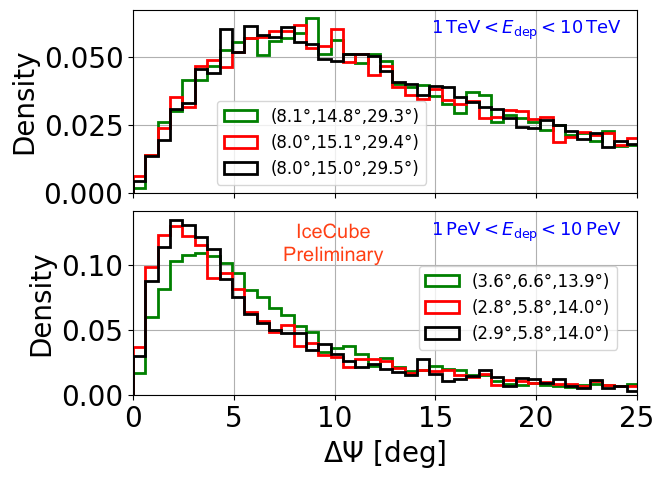}
    \end{subfigure}
    \caption{Left: Reconstruction resolution $\Delta\Psi$ on DNN Cascade MC sample under nominal, weaker, and stronger hole-ice assumptions, shown as a function of deposited cascade energy $E_{\text{dep}}$. Right: Distributions of $\Delta\Psi$ for the three scenarios in two energy slices. The legend indicates the (25\%, 50\%, 75\%) quantile values.}
    \label{DNN_resolutions_forward}
\end{figure}

\indent We also evaluate the impact of hole-ice knowledge bias on the DNN Cascade dataset which includes partially contained cascades. As expected, the angular resolutions (Fig.~\ref{DNN_resolutions_forward}, left panel) are worse than in the HESE sample which only contains contained showers. Below 100 TeV, hole-ice mismodeling has a negligible effect on reconstruction performance. In the PeV range, overestimating the absorption due to hole ice causes a modest median degradation of $\sim$1$^\circ$. This suggests that at lower deposited energies, resolution is largely driven by photon statistics rather than by systematic modeling errors. As the deposited energy increases and the photon statistics improve, the effects of hole-ice mismodeling become more pronounced. The resolution uncertainties, shown in the right panel of Fig.~\ref{DNN_resolutions_forward}, decrease slightly by about 1$^\circ$ for the nominal scenario compared to the overestimated hole-ice model in the 1–10 PeV energy range. Below 10 TeV, no improvements in uncertainties are observed with systematic mismodeling. Overall, the results suggest that with the anticipated improvements in modeling from the Upgrade, gains in reconstruction resolution for cascades are minimal. For high-energy tracks, the impact is expected to be even smaller.

\vspace{-1em} 

\begin{figure}[htbp]
    \centering
    \begin{subfigure}[t]{0.48\textwidth}
        \centering
        \includegraphics[width=\textwidth]{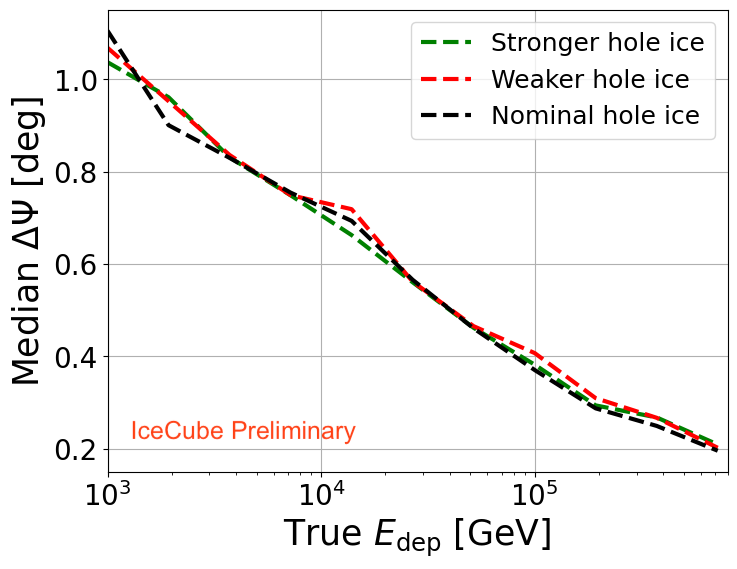}
    \end{subfigure}
    \hfill
    \begin{subfigure}[t]{0.48\textwidth}
        \centering
        \includegraphics[width=\textwidth]{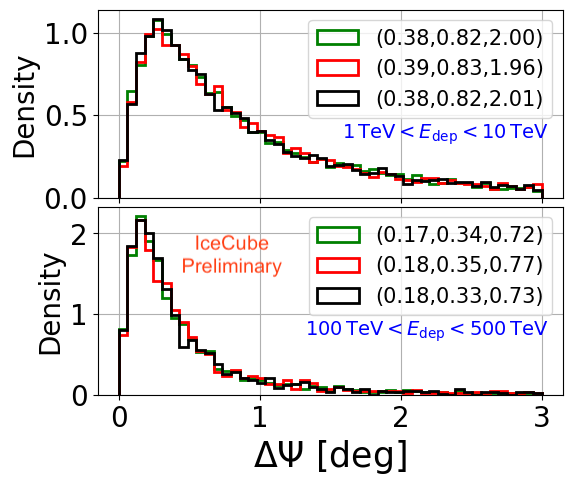}
    \end{subfigure}
    \caption{Left: Reconstruction resolution $\Delta\Psi$ on ESTES MC sample under nominal, weaker, and stronger hole-ice assumptions, shown as a function of deposited cascade energy $E_{\text{dep}}$. Right: Distributions of $\Delta\Psi$ for the three scenarios in two energy slices. The legend indicates the (25\%, 50\%, 75\%) quantile values.}
    \label{ESTES_resolutions_forward}
\end{figure}

\indent The resolution for high-energy muon tracks (Fig.~\ref{ESTES_resolutions_forward}, left panel) is an order of magnitude better than those for cascades due to the extended lever arm of the muon track which helps in a better directional estimate. This compensates for any resolution degradation from hole-ice mismodeling, resulting in minimal observable impact across the entire energy range. The right panel demonstrates that reductions in resolution uncertainties from improved systematic modeling are also negligible. Considering all track and cascade datasets, we conclude that the impact of hole-ice systematics on high-energy event reconstruction will be negligible after the Upgrade.
\subsection{Reconstruction bias}
\begin{figure}[htbp]
    \centering
    \begin{subfigure}[t]{0.32\textwidth}
        \centering
        \includegraphics[width=\textwidth]{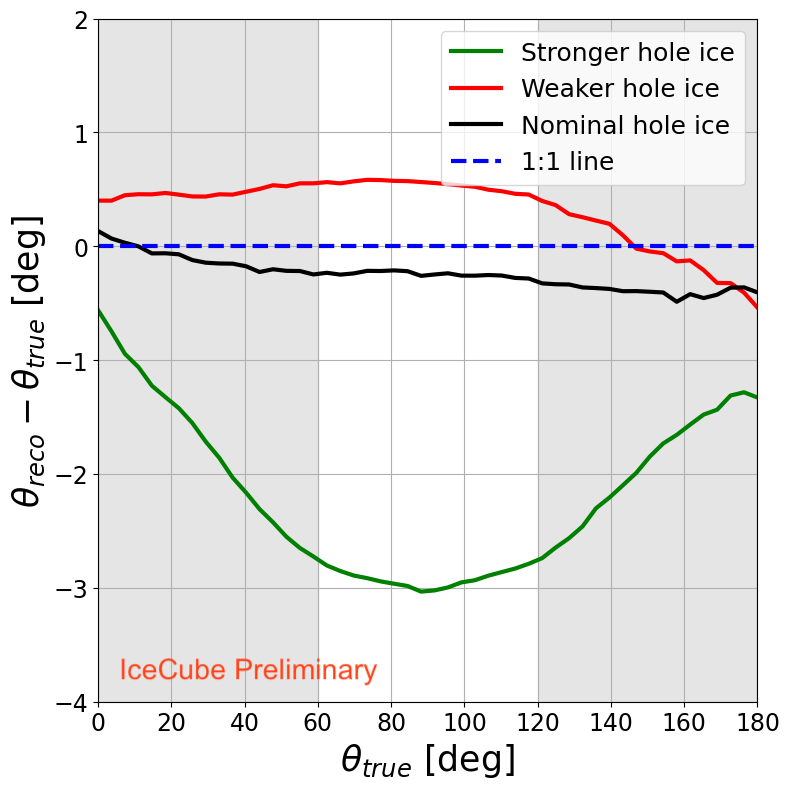}
    \end{subfigure}
    \hfill
    \begin{subfigure}[t]{0.32\textwidth}
        \centering
        \includegraphics[width=\textwidth]{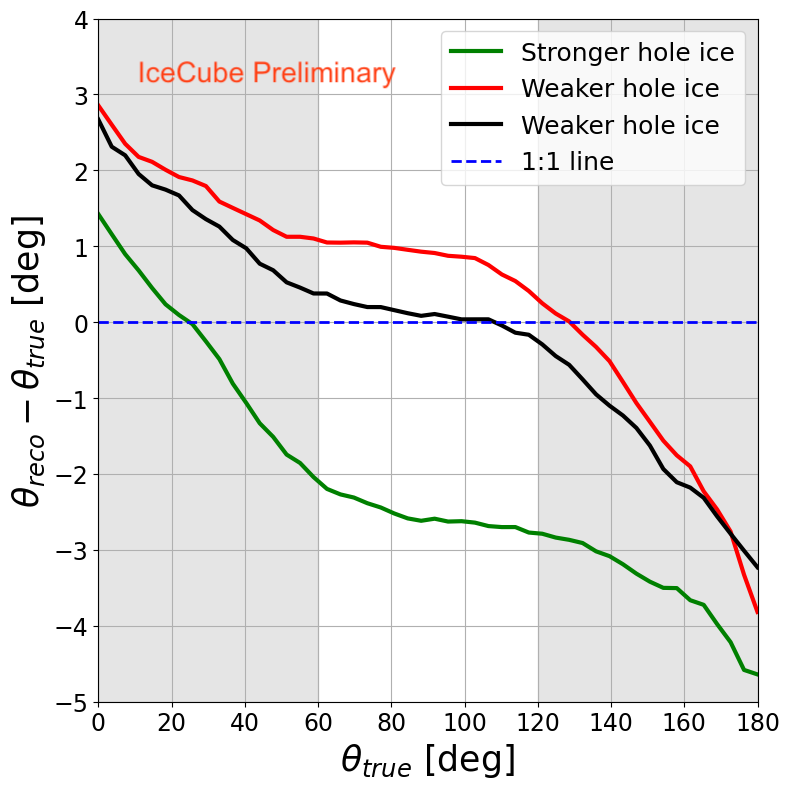}
    \end{subfigure}
    \hfill
    \begin{subfigure}[t]{0.32\textwidth}
        \centering
        \includegraphics[width=\textwidth]{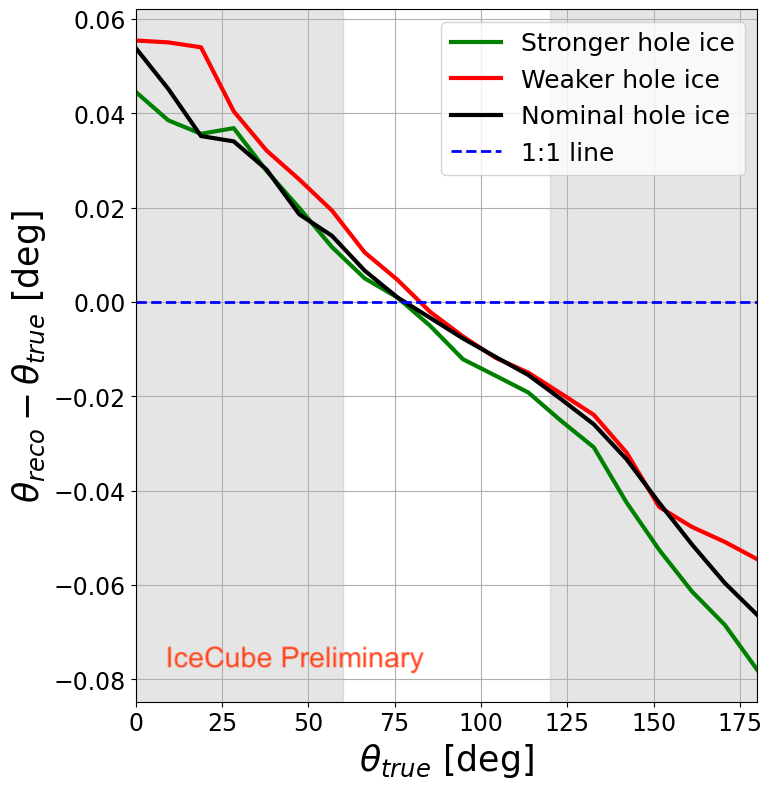}
    \end{subfigure}
    \caption{Comparison of median reconstruction biases across the HESE, DNN, and ESTES selections. The shaded regions indicate zenith angles where geometric constraints in the reconstruction introduce bias.}
    \label{fig:reco_bias_comparison}
\end{figure}

\indent Mis-modeling angular acceptance curves causes reconstruction bias by shifting the expected charge in hit DOMs which results in deviations from the true direction. For vertical events, reconstruction is biased both by hole-ice systematics and by a geometric limitation arising from the definition of the zenith angle: for near-vertical events (i.e., $\cos\theta_\text{true} \approx \pm1$), the reconstructed $\cos\theta$ can only deviate toward smaller absolute values, leading to an artificial bias. For horizontal events, the observed bias arises purely from hole-ice mis-modeling. In the HESE (Fig.~\ref{fig:reco_bias_comparison}, left panel) and DNN Cascade (Fig.~\ref{fig:reco_bias_comparison}, middle panel) datasets, a stronger suppression of the forward acceptance relative to the benchmark results in biases of up to 3$^\circ$.  The negative sign reflects a shift toward more downward-going reconstructions caused by reduced sensitivity to photons arriving at the front of the PMT. The bias for high-energy tracks (Fig.~\ref{fig:reco_bias_comparison}, right panel) is two orders of magnitude smaller than the cascade events. It is negligible for horizontal events for all cases of mismodelings. Overall, the results suggest that the bias will reduce by $\sim$1$^\circ$ over the existing systematic fits.

\vspace{-1em} 

\subsection{Source Sensitivities}
\indent The impact of uncertainties in hole ice parameters on reconstruction was translated into changes in the sensitivities of different source analyses. For both the HESE and DNN Cascade event selection, 90\% sensitivity and 5$\sigma$ discovery contours were evaluated under various hypotheses, including point sources and Galactic plane emission. This was done by injecting simulations based on the emission models and performing the full analysis pipeline. Metrics were compared between nominal and modified hole ice scenarios to evaluate the potential gains from improved angular resolution. For the Galactic plane hypothesis with 10 years of data, sensitivity improvements of $\sim$15\% and $\sim$10\% for the HESE and DNN samples, respectively, were observed. The effect on the discovery potential was found to be even smaller. This is a relatively minor improvement when compared with expectations for more data but can be applied to the dataset as a whole as ice knowledge improves.

\section{Summary \& Outlook}\label{sec4}
The IceCube Upgrade is expected to reduce hole ice parameter uncertainties to the sub-percent level. We find that the changes in the forward region of the modified acceptance curve influence reconstruction performance more strongly than changes in the lateral angular acceptance region. A weaker hole ice model, compared to the nominal case, shows a negligible impact across all datasets. For contained cascades, a stronger hole ice assumption results in a resolution degradation of $\sim$4$^\circ$ at 10 TeV and $\sim$1$^\circ$ for deposited energies above 100 TeV. No significant effect is seen for high-energy muon tracks. For horizontal cascades, the directional bias improves by up to $\sim$3°, while the effect on tracks is insignificant and largely independent of energy.\\
\indent The point source sensitivity study using the high-energy muon sample is still pending. Although substantial gains are not anticipated, a modest improvement over the diffuse source sensitivity is expected. A more comprehensive study could involve generating dedicated MC simulation sets by varying systematics such as DOM positions, individual module efficiency fluctuations, and scattering properties, and evaluating the impact of mis-modeling on angular resolution.

\bibliographystyle{ICRC}
\bibliography{references}
\clearpage

\section*{Full Author List: IceCube Collaboration}

\scriptsize
\noindent
R. Abbasi$^{16}$,
M. Ackermann$^{63}$,
J. Adams$^{17}$,
S. K. Agarwalla$^{39,\: {\rm a}}$,
J. A. Aguilar$^{10}$,
M. Ahlers$^{21}$,
J.M. Alameddine$^{22}$,
S. Ali$^{35}$,
N. M. Amin$^{43}$,
K. Andeen$^{41}$
% ...
\\
\\
% ...
$^{10}$ Universit{\'e} Libre de Bruxelles, Science Faculty CP230, B-1050 Brussels, Belgium \\
% ...
$^{16}$ Department of Physics, Loyola University Chicago, Chicago, IL 60660, USA \\
$^{17}$ Dept. of Physics and Astronomy, University of Canterbury, Private Bag 4800, Christchurch, New Zealand \\
% ...
$^{21}$ Niels Bohr Institute, University of Copenhagen, DK-2100 Copenhagen, Denmark \\
$^{22}$ Dept. of Physics, TU Dortmund University, D-44221 Dortmund, Germany \\
% ...
$^{35}$ Dept. of Physics and Astronomy, University of Kansas, Lawrence, KS 66045, USA \\
% ...
$^{39}$ Dept. of Physics and Wisconsin IceCube Particle Astrophysics Center, University of Wisconsin{\textemdash}Madison, Madison, WI 53706, USA \\
% ...
$^{41}$ Department of Physics, Marquette University, Milwaukee, WI 53201, USA \\
% ...
$^{43}$ Bartol Research Institute and Dept. of Physics and Astronomy, University of Delaware, Newark, DE 19716, USA \\
% ...
$^{63}$ Deutsches Elektronen-Synchrotron DESY, Platanenallee 6, D-15738 Zeuthen, Germany \\
$^{\rm a}$ also at Institute of Physics, Sachivalaya Marg, Sainik School Post, Bhubaneswar 751005, India \\
% ...

\subsection*{Acknowledgments}

\noindent
The authors gratefully acknowledge the support from the following agencies and institutions:
USA {\textendash} U.S. National Science Foundation-Office of Polar Programs,
U.S. National Science Foundation-Physics Division,
U.S. National Science Foundation-EPSCoR,
U.S. National Science Foundation-Office of Advanced Cyberinfrastructure,
Wisconsin Alumni Research Foundation,
Center for High Throughput Computing (CHTC) at the University of Wisconsin{\textendash}Madison,
Open Science Grid (OSG),
Partnership to Advance Throughput Computing (PATh),
Advanced Cyberinfrastructure Coordination Ecosystem: Services {\&} Support (ACCESS),
Frontera and Ranch computing project at the Texas Advanced Computing Center,
U.S. Department of Energy-National Energy Research Scientific Computing Center,
Particle astrophysics research computing center at the University of Maryland,
Institute for Cyber-Enabled Research at Michigan State University,
Astroparticle physics computational facility at Marquette University,
NVIDIA Corporation,
and Google Cloud Platform;
Belgium {\textendash} Funds for Scientific Research (FRS-FNRS and FWO),
FWO Odysseus and Big Science programmes,
and Belgian Federal Science Policy Office (Belspo);
Germany {\textendash} Bundesministerium f{\"u}r Forschung, Technologie und Raumfahrt (BMFTR),
Deutsche Forschungsgemeinschaft (DFG),
Helmholtz Alliance for Astroparticle Physics (HAP),
Initiative and Networking Fund of the Helmholtz Association,
Deutsches Elektronen Synchrotron (DESY),
and High Performance Computing cluster of the RWTH Aachen;
Sweden {\textendash} Swedish Research Council,
Swedish Polar Research Secretariat,
Swedish National Infrastructure for Computing (SNIC),
and Knut and Alice Wallenberg Foundation;
European Union {\textendash} EGI Advanced Computing for research;
Australia {\textendash} Australian Research Council;
Canada {\textendash} Natural Sciences and Engineering Research Council of Canada,
Calcul Qu{\'e}bec, Compute Ontario, Canada Foundation for Innovation, WestGrid, and Digital Research Alliance of Canada;
Denmark {\textendash} Villum Fonden, Carlsberg Foundation, and European Commission;
New Zealand {\textendash} Marsden Fund;
Japan {\textendash} Japan Society for Promotion of Science (JSPS)
and Institute for Global Prominent Research (IGPR) of Chiba University;
Korea {\textendash} National Research Foundation of Korea (NRF);
Switzerland {\textendash} Swiss National Science Foundation (SNSF).

\end{document}